\documentclass[english,12pt]{iopart}
\usepackage[T1]{fontenc}
\usepackage{subfigure}
\usepackage[latin1]{inputenc}
\usepackage{graphicx}
\usepackage{epstopdf}
\usepackage{epsfig}
\usepackage{prettyref}
\usepackage{babel}
\usepackage{color}
\makeatother

\begin{document}

\title[Coexistence of interacting-ferromagnetic and small-antiferromagnetic clusters]{Coexistence of interacting-ferromagnetic and small-antiferromagnetic clusters in La$_{0.5}$Ba$_{0.5}$CoO$_3$}

\author{Devendra Kumar and A Banerjee}
\address{UGC-DAE Consortium for Scientific Research, University Campus, Khandwa Road, Indore-452001,India}
\ead{deven@csr.res.in}

\begin{abstract}
We report detailed dc magnetization, linear and non-linear ac susceptibility measurements on the hole doped disordered cobaltite La$_{0.5}$Ba$_{0.5}$CoO$_3$. Our results show that the magnetically ordered state of the system consists of coexisting non-ferromagnetic phases along with percolating ferromagnetic-clusters. The percolating ferromagnetic-clusters possibly start a 
magnetic ordering at the Curie temperature of 201.5(5)~K. The non-ferromagnetic phases mainly consist of antiferromagnetic-clusters with size smaller than the ferromagnetic-clusters. Below Curie temperature the system exhibits an irreversibility in the field cooled and zero field cooled magnetization and frequency dependence in the peak of ac susceptibility. These dynamical features indicate towards the possible coexistence of spin-glass phase along with ferromagnetic-clusters similar to  La$_{1-x}$Sr$_{x}$CoO$_3$ (x$\geq$0.18), but the absence of field divergence in third harmonic of ac susceptibility and zero field cooled memory clearly rule out any such possibility. We argue that the spin-glass phase in La$_{1-x}$Sr$_{x}$CoO$_3$ (x$\geq$0.18) is associated with the presence of incommensurate antiferromagnetic ordering in non-ferromagnetic phases which is absent in La$_{0.5}$Ba$_{0.5}$CoO$_3$. Our analysis show that the observed dynamical features in La$_{0.5}$Ba$_{0.5}$CoO$_3$ are possibly due to progressive thermal blocking of ferromagnetic-clusters which is further confirmed by the Wohlfarth's model of superparamagnetism. The frequency dependence of the peak of ac susceptibility obeys the Vogel-Fulcher law with $\tau_0\approx10^{-9}$s. This together with the existence of an AT line in H-T space indicates the presence of significant inter-cluster interaction among these ferromagnetic-clusters.
\end{abstract}



\maketitle

\section{Introduction}
The transition metal oxides e.g. manganites, cuprates, and cobaltites exhibit complex phase diagram including the microscopically inhomogeneous electronic states due to interplay of various competitive electronic energies such as electron kinetic energy, electron-electron coulomb repulsion, spin-spin, spin-orbit, and crystal field interactions.\cite{Dagotto, Rini} Of these oxides, the cobaltite LaCoO$_3$ exhibits a unique property of temperature and doping dependent spin state transition.\cite{Louca1, Podlesnyak} The Co$^{3+}$ ion in LaCoO$_3$ can exist in low spin (LS) state with configuration t$_{2g}^6$e$_g^0$ ($S$=0), intermediate spin (IS) state with configuration t$_{2g}^5$e$_g^1$ ($S$=1), and high spin (HS) state with configuration t$_{2g}^4$e$_g^2$ ($S$=2). The LaCoO$_3$ have a charge transfer insulator type non-magnetic  ground state with Co$^{3+}$ ion in the LS state; it starts showing magnetic moment above 30~K and exhibits a paramagnetic like behavior above 100~K.\cite{Raccah, Kriener} This change in magnetic moment and behavior is attributed to thermally driven spin state transition of Co$^{3+}$ ion,  but the nature of transition whether it is a LS-IS transition or LS-HS transition is still not completely settled.\cite{Podlesnyak, Raccah, Kriener, Zobel, Noguchi, Haverkort, Radaelli, Korotin} The hole doping of LaCoO$_3$ by replacing the trivalent La$^{3+}$ with divalent Sr$^{2+}$ or Ba$^{2+}$ generates Co$^{4+}$, and each of these Co$^{4+}$ transforms their six nearest Co$^{3+}$ neighbors into the IS state by forming octahedrally shaped spin-state polarons.\cite{Louca, Phelan, Podlesnyak1} In these polarons, e$_g$ electrons of Co$^{3+}$ are delocalized and are shared by Co$^{3+}$ and Co$^{4+}$ ions of the polaron, while t$_{2g}$ electrons of both the ions are localized and couple ferromagnetically via double exchange interaction. For small hole doping
these isolated spin state polarons are stable within the nonferromagnetic matrix. Additional hole doping enhances the number density of spin state polarons, and above a critical doping of x=0.04, the enhanced polaron density  causes a decay of polaronic state due to ferromagnetic (FM) interaction between the intra-polaronic Co$^{3+}$ ions at the cost of the antiferromagnetic (AFM) intra-polaronic interaction.\cite{Podlesnyak2} This, in turn, results in the formation of hole rich ferromagnetic spin clusters embedded in non-ferromagnetic insulating matrix. On further enhancing the hole doping, at a critical concentration ($x$=0.18 for Sr$^{2+}$ and $x$=0.2 for Ba$^{2+}$), the ferromagnetic metallic clusters eventually percolates giving rise to long range ferromagnetic ordering and metallic conductivity.\cite{Kriener1}

For La$_{1-x}$Sr$_{x}$CoO$_3$, a spin-glass state is observed below the critical doping concentration for percolation of ferromagnetic metallic regions and above this a ferromagnetic or a ferromagnetic-cluster state is reported.\cite{Kriener1, Wu}
Detail investigations of ferromagnetic-cluster state of  La$_{1-x}$Sr$_{x}$CoO$_3$ suggest the presence of spin or cluster -glass like behavior even in the so called ferromagnetic or ferromagnetic-cluster state\cite{Tang, Mukherjee1, Samal, Samal1, Samal2} and it has been been argued that this behavior is due to coexistence of the spin-glass phase along with  percolating ferromagnetic-clusters.\cite{Tang, Samal1} The absence of exchange bias effect in La$_{0.5}$Sr$_{0.5}$CoO$_3$ clearly indicates that the spin-glass like phase is not present at the interface of ferromagnetic-clusters and non-ferromagnetic matrix, but instead, it probably coexist as small patches along with the percolating backbone of ferromagnetic-clusters.\cite{Samal1} The nature of magnetic state in La$_{1-x}$Ba$_{x}$CoO$_3$ with Ba$^{2+}$ having a larger ionic radii than Sr$^{2+}$ (ionic radii of La$^{3+}$=1.216\AA, Ba$^{2+}$=1.47\AA, Sr$^{2+}$=1.31\AA) is relativity less studied, and early reports indicate the presence of ferromagnetic-metallic ground state for $x>0.2$.\cite{Kriener1, Mandal} The higher ionic radii of Ba$^{2+}$ (a) enhances the local randomness due to larger size mismatch between the Ba$^{2+}$ and La$^{3+}$ ions, and (b) reduces the overall distortion from ideal pervoskite structure and so the tolerance factor ($t$) approaches to 1. This enhancement in tolerance factor straightens the Co-O-Co bonds which in turn increases the ferromagnetic coupling due to double exchange interaction between Co$^{3+}$  and Co$^{4+}$  ions. Furthermore, the Ba$^{2+}$ doping enhances the concentration of Jahn-Teller (J-T) active IS state because of lattice expansion and the formation of J-T magnetopolaron is found to be most preferable in the insulating phase of Ba doped cobaltites.\cite{Phelan, Phelan1}

In this paper we present results of detailed dc magnetization, linear, and nonlinear ac susceptibility measurements of La$_{0.5}$Ba$_{0.5}$CoO$_3$ with an aim to understand its magnetically ordered state. We find that the magnetically ordered state of La$_{0.5}$Ba$_{0.5}$CoO$_3$ consists of small antiferromagnetic-clusters coexisting along with the percolating backbone of ferromagnetic-clusters. In contrast to La$_{0.5}$Sr$_{0.5}$CoO$_3$ no signatures of spin-glass phase have been observed in La$_{0.5}$Ba$_{0.5}$CoO$_3$. Our analysis suggests that the existence of spin-glass phase in hole doped LaCoO$_3$ (above critical concentration) is associated with the presence of incommensurate antiferromagnetic ordering in the non-ferromagnetic phases which in turn depends on the ionic radii and doping level of divalent ion. Furthermore we show that the observed dynamic properties in La$_{0.5}$Ba$_{0.5}$CoO$_3$ comes from the progressive thermal blocking of interacting ferromagnetic-clusters.

\section{Experimental Details}
Polycrystalline La$_{0.5}$Ba$_{0.5}$CoO$_3$ samples are prepared by pyrophoric method~\cite{Pati} using high purity (99.99\%) La$_2$O$_3$, BaCoO$_3$, and Co(NO$_3$)$_2$6H$_2$O. The stichometric ratio of La$_2$O$_3$, BaCoO$_3$, and Co(NO$_3$)$_2$6H$_2$O are separately dissolved in dilute nitric acid and then these solutionis are mixed with the triethanolamine (TEA) keeping the pH highly acidic. The final solution is dried at ~100~$^\circ$C, which burns and yields a black powder that is palletized and subsequently annealed at 1100~$^\circ$C for 12~hour. These samples are characterized by X-Ray diffraction on a Bruker D8 Advance X-ray diffractometer using Cu-K$\alpha$ radiation. The dc magnetization measurements are performed on a 14~T Quantum Design physical property measurement system-vibrating sample magnetometer and the low field ac susceptibility measurements are carried out on a ac-susceptibility setup which is described in reference~\cite{Bajpai}.

\begin{figure}[!t]
\begin{centering}
\includegraphics[width=0.8\columnwidth]{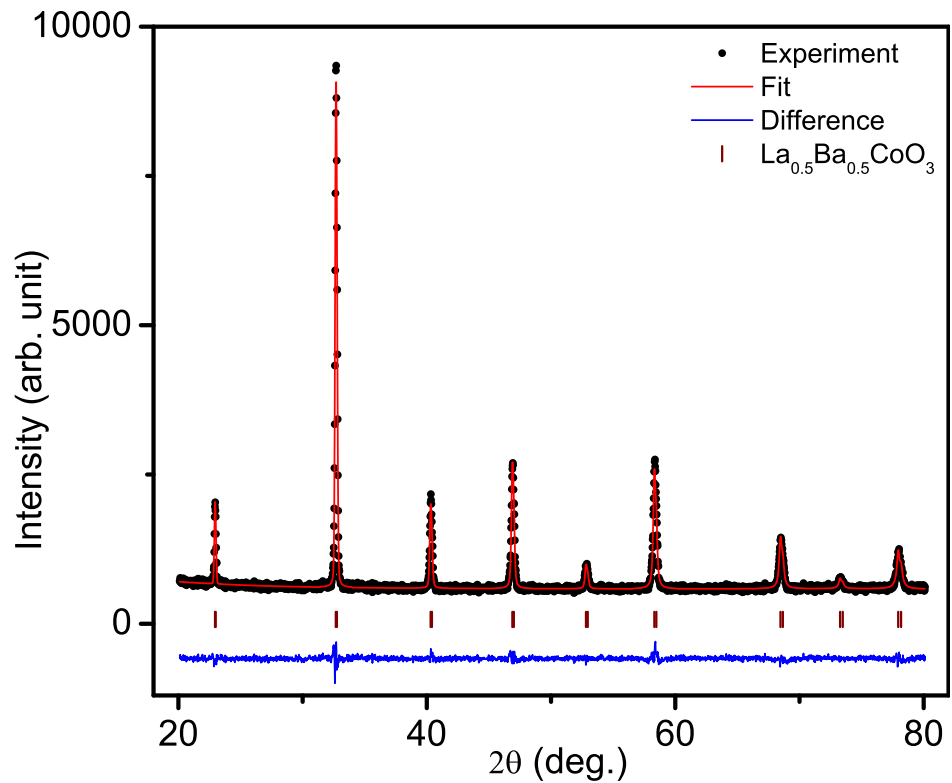}
\par\end{centering}
\caption{Room temperature X-ray diffraction pattern of La$_{0.5}$Ba$_{0.5}$CoO$_3$. The solid circles show the experimental X-ray diffraction data, the red line on the experimental data shows the Rietveld refinement for simple cubic Pm-3m structure with $\chi^2$=1.34, the short vertical lines give the Bragg peak positions, and the bottom blue line gives the difference between the experimental and calculated pattern.} \label{fig: XRD}
\end{figure}

The X-ray diffraction data of La$_{0.5}$Ba$_{0.5}$CoO$_3$ is collected at room temperature and analyzed with Rietveld structural refinement using FULLPROF software.\cite{Carvajal} Figure \ref{fig: XRD} shows the XRD data, the Rietveld fit profile, the Bragg positions, and the difference in experimental and model results. The Rietveld refinement show that the sample is single phase and crystallizes in simple cubic Pm-3m structure with lattice constant $a$=3.8726(2)\AA{} and  unit cell volume $V$=58.078(4)\AA$^3$. The unit cell volume in disordered cobaltites depends on the oxygen stoichiometry, and the comparison of  our result with that of reference~\cite{Troyanchuk} suggests that the oxygen non-stoichiometry ($\delta$ in La$_{0.5}$Ba$_{0.5}$CoO$_{3+\delta}$) is much less than 0.05. 
The oxygen content is determined by iodometric titration which gives $\delta$=0.00(2). The fluctuation in the average atomic concentration of La, Ba, and Co is probed by energy dispersive analysis of x-ray (EDAX) attached with TECNAI~G2-20FEI transmission electron microscope. EDAX measurements at the step of 0.5~$\mu$m along a randomly chosen straight line of 5~$\mu$m length give a variation of less than 0.5~\% in the average atomic concentration, which is within the experimental uncertainty ($\sim$1\%) of EDAX. This shows that the La and Ba are uniformity distributed in the sample. The average crystallite size of the sample is estimated from XRD data using the Scherrer formula which comes around 85~nm.

\section{Results and Discussions}
\subsection{DC Magnetization}
\subsubsection{Thermomagnetic irreversibility}
\begin{figure}[!t]
\begin{centering}
\includegraphics[width=0.8\columnwidth]{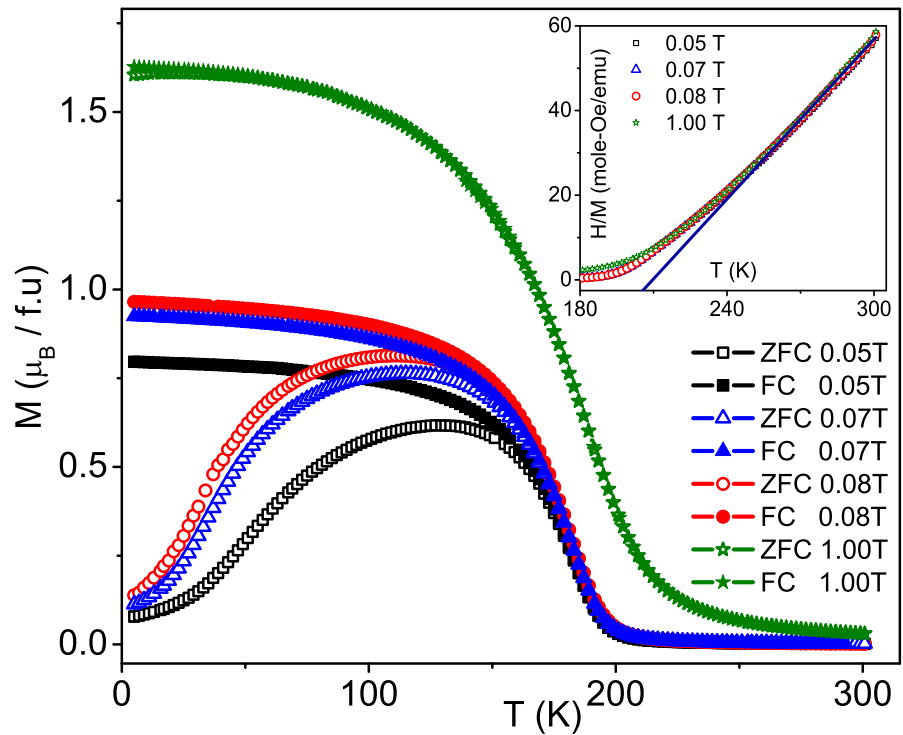}
\par\end{centering}
\caption{Temperature dependence of  magnetization under FC and ZFC protocol at various measuring fields. The inset shows the H/M versus temperature for 0.05, 0.07, 0.08, and 1.0~T; and the solid line is the Curie-Weiss fitting of 0.08~T curve above 250~K.} \label{fig: MvsT}
\end{figure}

Figure \ref{fig: MvsT} show the temperature variation of magnetization in field cooled (FC) and zero field cooled (ZFC) protocol. In the FC protocol, the sample is cooled to 5~K in presence of measuring field and the magnetization is recorded in heating run keeping the field constant. In ZFC protocol the sample is cooled to 5~K in zero field and then the measuring field is applied and magnetization is recorded as a function of temperature in the heating run. On lowering the temperature, around 200~K,  both the FC and ZFC magnetization curves show a rapid increase in magnetization which is indicative of paramagnetic to ferromagnetic transition. On further lowering the temperature, the FC curve keeps evolving while the ZFC curve bifurcates with that of FC at the temperature T$_{irr}$ and exhibits a broad peak at a temperature $T_p$. On increasing the measuring field $T_{irr}$ and $T_p$ decreases with an enhancement in broadening of ZFC peak. At 1~T, the FC and ZFC curves almost coincide. The bifurcation in FC-ZFC magnetization along with a peak in ZFC magnetization indicates about the presence of a spin-glass,\cite{Maydosh} cluster-glass,\cite{Deac, Huang} super-paramagnetic,\cite{Knobel, Pramanik} or anisotropic ferromagnetic state.\cite{Anil} At low fields, $T_p$ < $T_{irr}$, and below $T_p$ the FC magnetization is not constant with temperature. This observation is not in agreement with that of canonical spin-glasses and suggests that the system is possibly in a  cluster-glass, super-paramagnetic, or ferromagnetic state. Similar observations have been made on the other relatively well studied half doped disordered cobaltite La$_{0.5}$Sr$_{0.5}$CoO$_3$.\cite{Wu, Samal, Samal1}

The high temperatures magnetization data (T>250~K) fits well with Curie-Weiss law, and for 0.08~T filed cooling, Curie constant $C$=1.59(1)~emu-K/mole-Oe and Weiss constant $\theta$=210(2)~K. See inset of figure \ref{fig: MvsT}. The coefficients of fitting show a small but unsystematic field dependence with a variation of 0.08~emu-K/mole-Oe in $C$ and 3~K in $\theta$ in the field range of 0.05-1.0~T. The Curie constant of 1.59(1)~emu-K/mole-Oe gives an effective value of paramagnetic moment $\mu_{eff}=$ 3.566(3)~$\mu_B$/f.u. and the positive value of Weiss constant indicates the dominance of ferromagnetic correlations in the ordered state. Below 250~K, the inverse dc susceptibility exhibits a upward deviation from the Curie-Weiss law. A similar upward deviation from Curie-Weiss law above T$_C$ has also been observed in La$_{1-x}$Sr$_x$CoO$_3$ and it was attributed to the existence of short range ferromagnetic correlations above T$_C$.\cite{He}   
\subsubsection{Coexistence of ferromagnetic and non-ferromagnetic phases}
\label{subsubsection:isothermal magnetization}
\begin{figure} [!t]
\begin{centering}
\includegraphics[width=0.8\columnwidth]{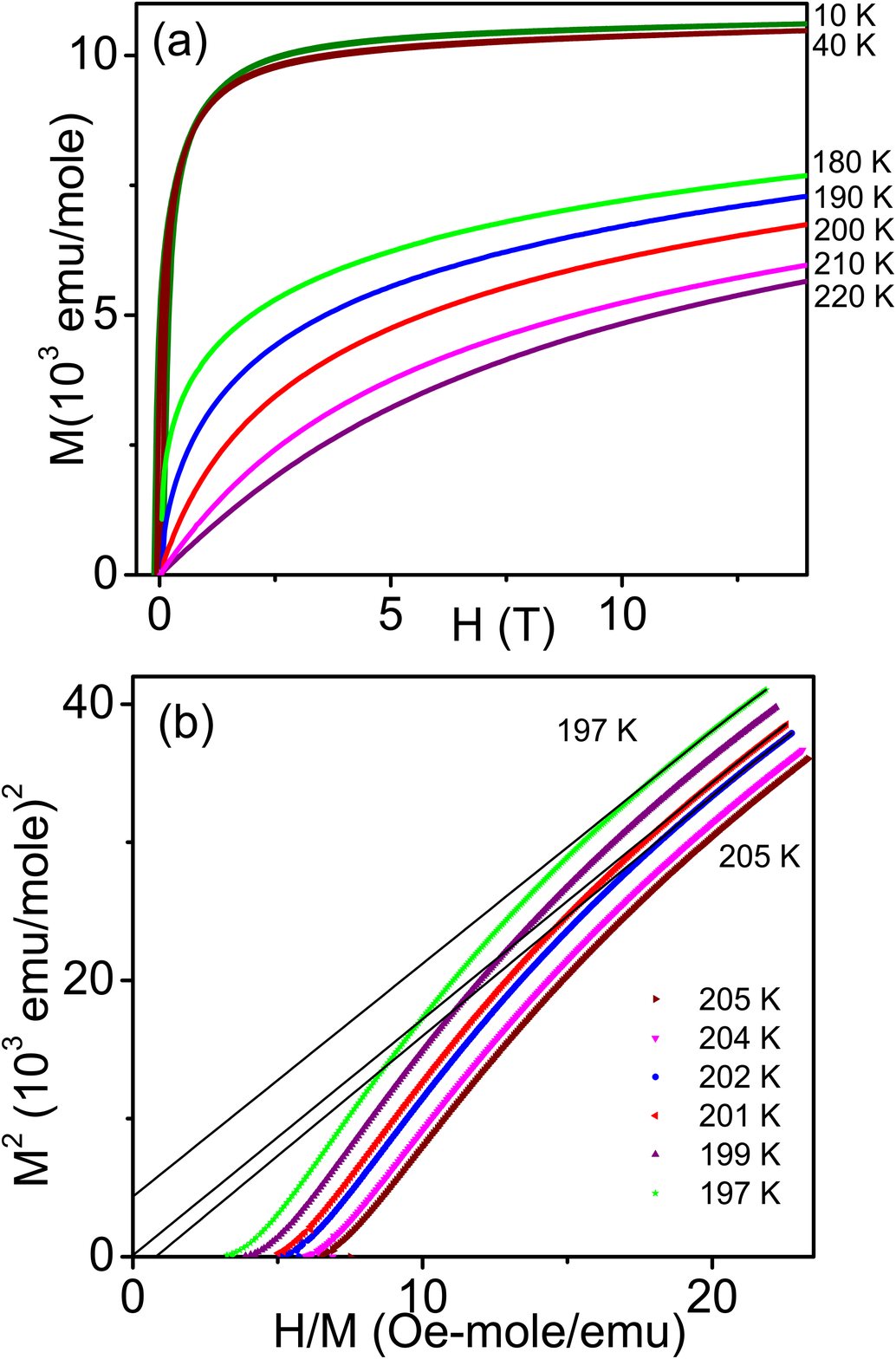}
\par\end{centering}
\caption{(a) Magnetization versus field at 10~K, 40~K, 180~K, 190~K, 200~K, 210~K, and 220~K (b) Arrot plot (M$^2$ vs. H/M) of the magnetization isotherms at 197~K, 199~K, 201~K, 202~K, 204~K, and 205~K. The solid black lines are straight line fit to M$^2$ vs. H/M curve at high field which are extrapolated to H=0.} \label{fig:MH}
\end{figure}

In figure \ref{fig:MH} (a) we show the magnetization versus field plot at 10~K, 40~K, 180~K, 190~K, 200~K, 210~K, and 220~K. At low temperatures, for example at 10~K the magnetization exhibits a saturation like behavior at high magnetic fields which is typical of a ferromagnet, but a careful observation of the data indicates the presence of a non-saturating magnetization along with the saturating ferromagnetic component. The presence of this non-saturating component prohibits the magnetization from saturating even at high magnetic fields. The existence of a non-ferromagnetic component along with the ferromagnetic component is in agreement with the cluster model of other disordered cobaltite La$_{1-x}$Sr$_x$CoO$_3$ where a number of studies have shown the presence of non-ferromagnetic Co$^{3+}$ phases that coexist along with ferromagnetic-clusters.\cite{Wu, Samal} The ferromagnetic component can be extracted from the total magnetization by assuming that the total magnetization can be written as $M_{tot}$=$M_F$ + $\chi_{AF}$H where $M_F$ is the saturation value of ferromagnetic component and $\chi_{AF}$ is the slope of M vs. H curve at high fields. Using this to fit the magnetization versus field curve above 12~T at 10~K, we estimate the saturation magnetization of ferromagnetic component as 1.855(1)~$\mu_B$/Co. The value of saturation magnetization of ferromagnetic-clusters is smaller than that expected from the spin only value ($M_s=gS\mu_B=2.5\mu_B$) when both the Co$^{3+}$ ($S$=1) and Co$^{4+}$ ($S$=3/2) are in IS state. It is to be noted that the similar results about the difference in experimental and expected saturation magnetization have also been reported on La$_{1-x}$Sr$_x$CoO$_3$.\cite{Wu, Samal} On the basis of the band structure calculations, Ravindra $et$~$al.$\cite{Ravindran} have shown that the hole doping in these materials reduces the ionicity, enhances the Co-O hybridization, and stabilizes the IS state. Due to enhanced Co-O hybridization the expected average Co moment is reduced compared to the prediction of simple ionic model.

In figure \ref{fig:MH} (b), the M-H isotherms around 200~K are plotted as M$^2$ versus H/M which is known as Arrot plot.\cite{Arrott} In these plots, the intercept of the linear fitting of high field data on the X and Y axis gives inverse susceptibility and spontaneous magnetization respectively and the one passing through origin gives the ferromagnetic transition temperature T$_C$. At 201~K, the value of spontaneous magnetization is 0.058~$\mu_B$/Co which shows the presence of ferromagnetic interactions, and  the line passing through origin will correspond to M$^2$ versus H/M curve lying in between 201-202~K indicating that the T$_C$ lies in between. The curves in the Arrot plot exhibit a downward curvature even at moderate fields. This suggests the possibility of non-mean-field like behavior and therefore modified Arrot plots are better suited for more accurate determination of T$_C$. Above T$_C$, e.g. at 204~K and 205~K, the spontaneous magnetization (M$_S$) is zero indicating the absence of long range ferromagnetic ordering.

\subsection{AC Susceptibility}
In order to get a better understanding of the magnetically ordered state, we have performed ac susceptibility measurements at low fields which probe the dynamics of the system at the time scales decided by the measuring frequency range. The magnetization ($M$) of a system can be expressed in terms of the applied field ($H$) as:
\begin{equation}
M(H)=M_0+\chi_1H+\chi_2H^2+\chi_3H^3+...\label{eq:acsusp}
\end{equation}
where $M_0$ is the spontaneous magnetization, $\chi_1$ is the linear susceptibility and $\chi_2$, $\chi_3$,.. are the nonlinear susceptibilities which can be identified with the Taylor series expansion of $M(H)=M_0+(1/1!)(dM/dH)_{H=0}H+(1/2!)(d^2M/dH^2)_{H=0}H^2+..$ .
\subsubsection{Nature of magnetically ordered state}
\label{subsubsection:Nature of magnetically ordered state}
\begin{figure} [!t]
\begin{centering}
\includegraphics[width=0.8\columnwidth]{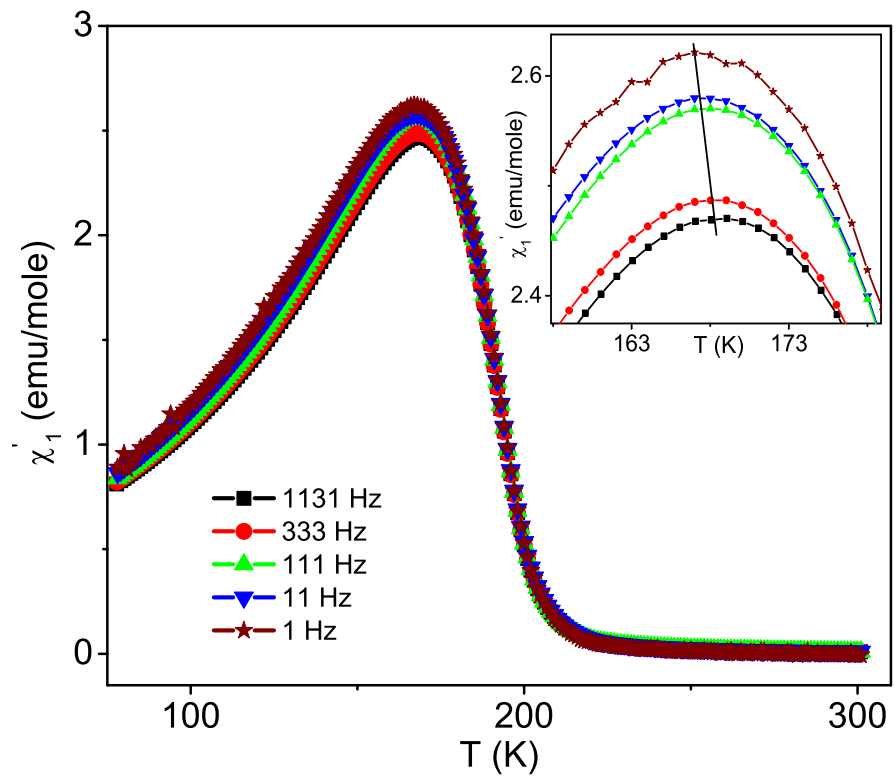}
\par\end{centering}
\caption{Temperature dependence of the real part of linear ac susceptibility at various frequencies at the ac field of 2.21~Oe. The inset shows the expanded view of the peak in ac susceptibility.} \label{fig: ACR}
\end{figure}

Figure \ref{fig: ACR} show the real part of linear ac susceptibility ($\chi^{'}_1$) measured in the ac field of 2.21~Oe and frequency 1131, 333, 131, 11, and 1~Hz. The $\chi^{'}_1$ exhibits a broad peak similar to that of ZFC magnetization and the peak position ($T_B$) in $\chi^{'}_1$ increases on increasing the measuring frequency ($\nu$) (see inset of figure \ref{fig: ACR}) which is a common feature of spin-glass, cluster-glass, and superparamagnetic systems; and the presence of frequency dependence in $T_B$ clearly rules out the possibility of normal long range ferromagnetic state. The frequency dependence in $\chi^{'}_1$ is quantified as $\Phi=\Delta T_B/(T_B\Delta$log$_{10}f)$ and the estimated value of $\Phi$ is 0.0023 which is in agreement with typical values seen in canonical spin-glasses, cluster-glasses, superspin-glasses, or interacting superparamagnets (0.02-0.005),\cite{Goya, Cong, Thakur} and two order of magnitude lower than that observed in noninteracting superparamagnets (0.1-0.3).\cite{Toro, Dormann}

\begin{figure} [!t]
\begin{centering}
\includegraphics[width=0.9\columnwidth]{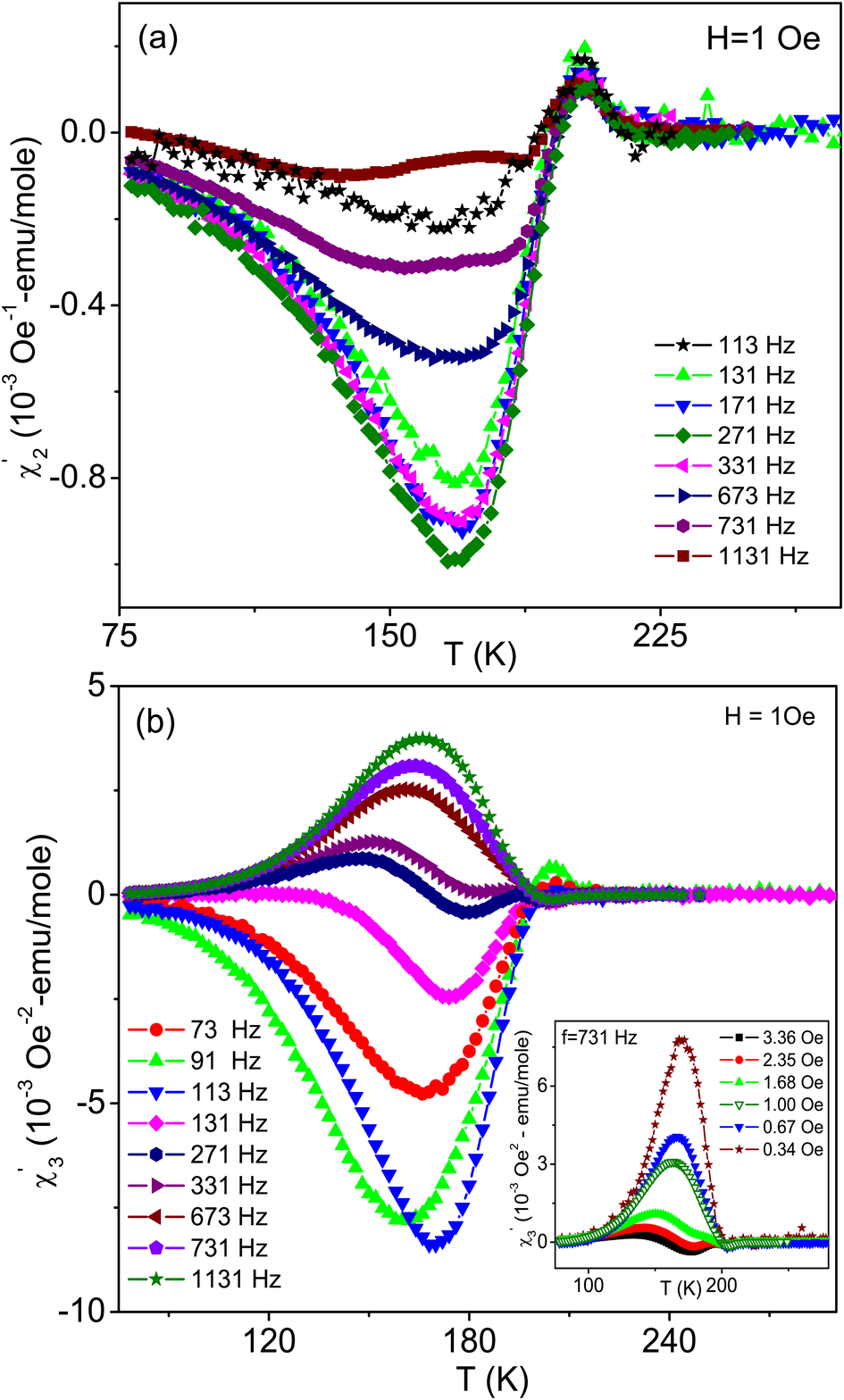}
\par\end{centering}
\caption{Temperature dependence of the real part of (a) second harmonic and (b) third harmonic  of ac susceptibility at various frequencies at the ac field of 1~Oe. The inset of figure (b) shows the temperature dependence of third harmonic of ac susceptibility at 731~Hz at various ac fields.} \label{fig: chi2}
\end{figure}

In absence of the time inversion symmetry breaking field, $M(H)$=-$M(-H)$, and all the even terms in equation \ref{eq:acsusp} i.e. $\chi_2$, $\chi_4$ are zero. The $\chi_2$ is observed in presence of a superimposed external dc field or an internal field which originates from magnetically correlated spins. For canonical spin-glass the coefficient of even powers of $H$ in equation \ref{eq:acsusp} are zero. The real part of nonlinear susceptibility $\chi^{'}_2$ is plotted in figure \ref{fig: chi2}~(a). $\chi^{'}_2$ is zero in paramagnetic phase, has a small positive peak at 202~K, then a large negative peak around 167~K, and thereafter it slowly approaches to zero. The strength of negative peaks in $\chi^{'}_2$ diminishes on increasing the ac frequency. Below T$_C$, the negative value of $\chi^{'}_2$ clearly show the presence of ferromagnetic ordering which also rules out the presence of canonical spin-glass state, but the possibility of coexistence of spin-glass phase along with ferromagnetic-clusters remains open. 

Figure \ref{fig: chi2}~(b) show the temperature dependence of the real part of third harmonic of ac susceptibility ($\chi^{'}_3$) at different measurement frequencies. At the lower frequencies $\chi^{'}_3$ exhibit a broad negative peak similar to superparamagnets or spin/cluster-glasses, on increasing the frequency $\chi^{'}_3$ changes sign similar to ferromagnets,\cite{Suzuki, FUJIKI} and on further increasing the frequency a positive peak in $\chi^{'}_3$ is observed. In the critical regime, the characteristic relaxation time ($\tau$) depends on the dynamic spin-spin correlation length ($\xi$) as $\tau\propto\xi^{z}$ where $z$ is the dynamic critical exponent. On increasing the measurement frequency ($\nu$), the times scale of relaxations that can be probed through ac susceptibility measurement decreases, and therefore, the accessible region of relaxations in the ac susceptibility measurements shift towards smaller $\tau$ and $\xi$. 
When accessible $\xi$ reduces to the length scale of ferromagnetic-clusters, then contribution from within ferromagnetic-clusters dominates $\chi^{'}_3$ and we get a ferromagnetic like critical behavior in $\chi^{'}_3$. At 673~Hz and above $\chi^{'}_3$ exhibits a positive peak. The strength of this peak increases on (a) increasing the measurement frequency and (b) lowering the ac field (see inset of figure \ref{fig: chi2}~(b)). For antiferromagnets with coordination number $n\leq6$, in the framework of Bethe approximation, $\chi_3$ is always positive. It grows as temperature increases towards Neel temperature (T$_N$) and exhibits a discontinuity at T$_N$.\cite{FUJIKI} Experimentally a positive peak in $\chi^{'}_3$ has been observed in an antiferromagnet with $n\leq6$.\cite{Cimberle} This suggests that the observed positive peak in $\chi^{'}_3$ at high ac frequencies is due to presence of antiferromagnetic-clusters. 
The average size of these antiferromagnetic-clusters is expected to be smaller than that of ferromagnetic-clusters.  This is because at higher ac frequencies, the contribution from regions of smaller relaxation time (and so smaller correlation length) determines the overall behavior of $\chi^{'}_3$.
On the basis of this, we infer that the non-ferromagnetic phases discussed in section \ref{subsubsection:isothermal magnetization} mainly consist of small antiferromagnetic-clusters (average size smaller than that of ferromagnetic-clusters) that coexist along with percolating backbone of ferromagnetic-clusters.

\begin{figure} [!t]
\begin{centering}
\includegraphics[width=0.8\columnwidth]{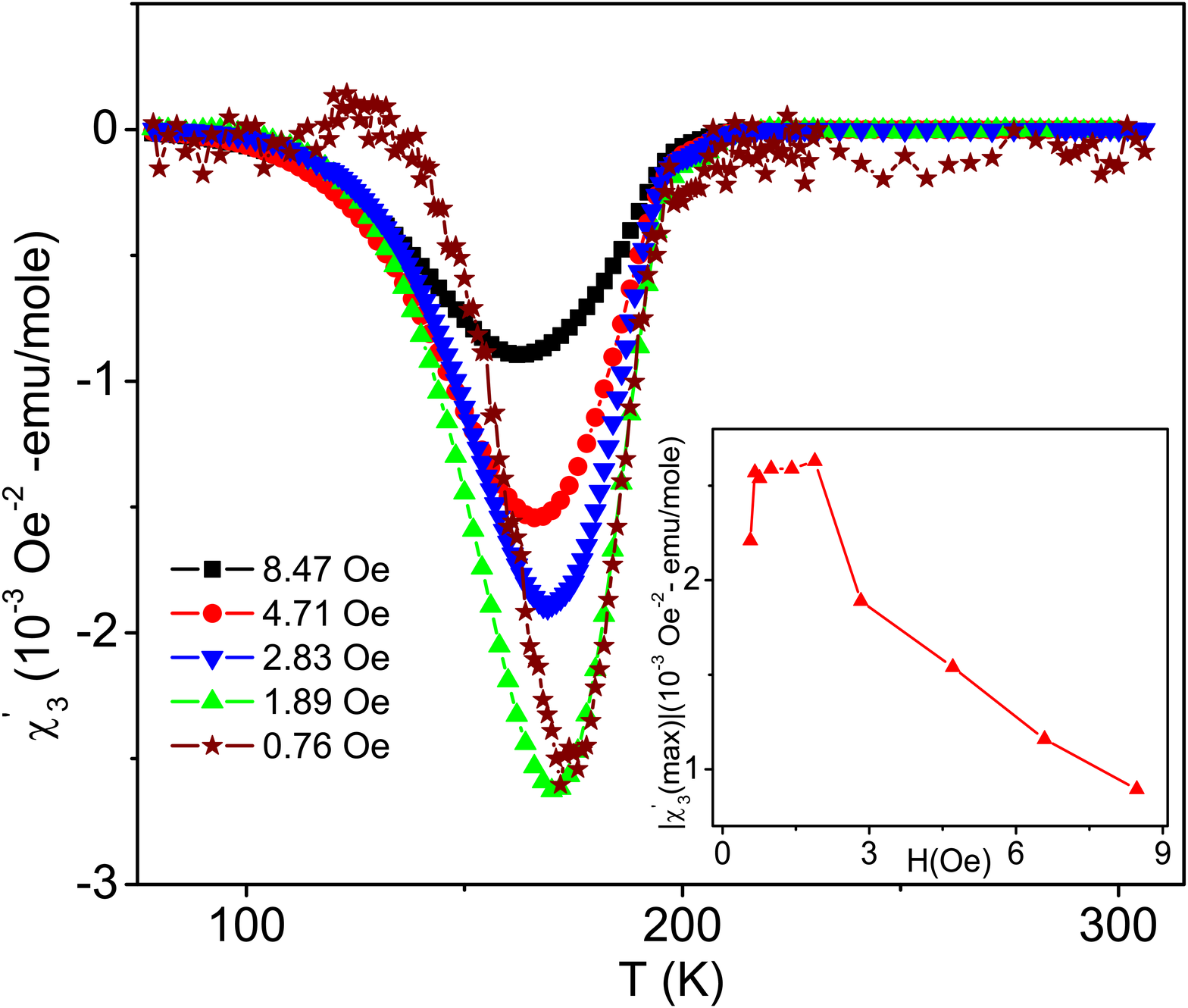}
\par\end{centering}
\caption{Temperature dependence of the real part of third harmonic of ac susceptibility at various ac fields at 131~Hz. The inset shows the ac field dependence of the peak value of $\chi_{3}$ which is $\chi_{3}^{'}(max)$.} \label{fig: chi3}
\end{figure}

\subsubsection{Absence of spin or cluster -glass like transition}
After discarding the possibility of canonical spin-glass state on the basis of non zero value of $\chi_2$
we need to identify the origin of frequency dependence in $T_B$ from remaining possibilities, which are the cluster-glass, the super-paramagnetism, and the coexistence of spin-glass phase along with ferromagnetic-clusters. In the first two cases, relaxing entities are super-spins i.e. the moment of a single magnetic domain (cluster); 
while for the last case the relaxing entities are  the atomic-spins. It is quite difficult to distinguish whether the slowing down in spin dynamics is due to progressive thermal blocking or due to spin-glass like cooperative freezing of the fluctuating entities.
To determine the nature of spin dynamics, we have measured the third harmonic of ac susceptibility ($\chi_3$) which is proportional (and opposite in sign) to spin-glass susceptibility ($\chi_{SG}$). The negative divergence of $\chi_3$ at $T_g$ in the limit of $H\to0$ gives the direct evidence of spin-glass like critical slowing down of the fluctuating entities and hence can be used as an unambiguous test to probe the presence of spin or cluster -glass phase.\cite{Suzuki, Chikazawa, Bajpai1} The temperature dependence of the real part of the third harmonic of ac susceptibility at 131~Hz and at different ac fields is plotted in figure~\ref{fig: chi3}. The magnitude of the peak in $\chi^{'}_3$ ($\chi_{3}^{'}$(max)) depends on the ac field and the field dependence of $\chi_{3}^{'}$(max) is plotted in the inset of figure \ref{fig: chi3}. The $\chi_{3}^{'}$(max) does not diverge as $H\to0$ which clearly shows that the fluctuating entities do not freeze in a spin or cluster -glass state. We do not observe any ZFC memory effect which further supports the absence of spin or cluster -glass like freezing in the system.
These results suggest that the observed frequency dependence in $\chi^{'}_1$ is possibility due to progressive thermal blocking of fluctuating entities.

\begin{figure} [!t]
\begin{centering}
\includegraphics[width=0.8\columnwidth]{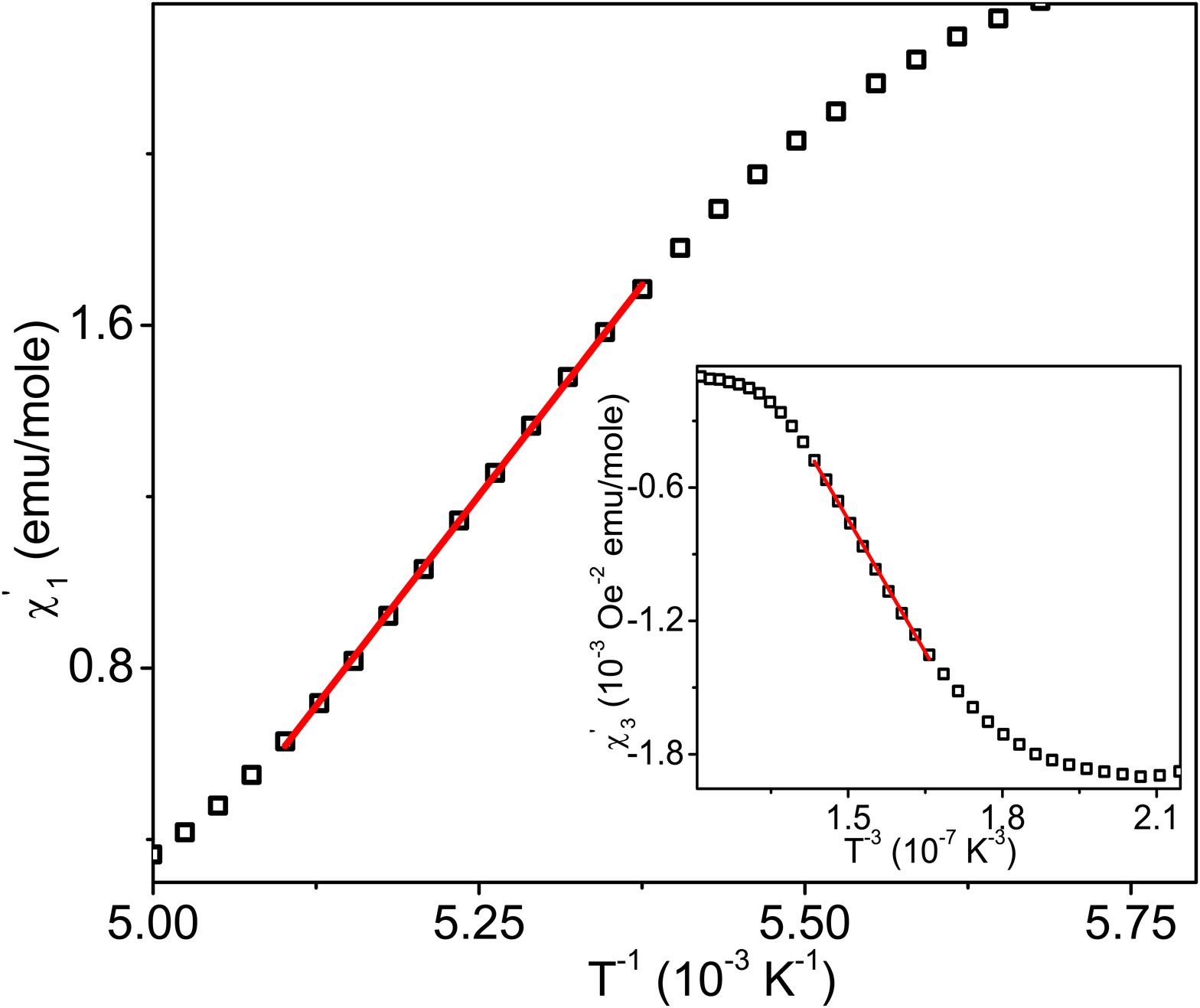}
\par\end{centering}
\caption{$\chi^{'}_1$ versus $T^{-1}$ above the blocking temperature. The inset shows the $\chi^{'}_3$ versus $T^{-3}$ above the blocking temperature. The straight lines are the fitting of equation \ref{eq:Wohlfarth1} and \ref{eq:Wohlfarth2} to the data.} \label{fig: superpara}
\end{figure}

\subsubsection{Superparamagnetic behavior of ferromagnetic clusters}
The existence of superparamagnetic behavior, i.e. progressively thermal blocking of single domain magnetic clusters is further substantiated by Wohlfarth's model of superparamagnets  which shows that the magnetization of an ensemble of magnetic clusters is given as\cite{Wohlfarth, Bitoh}
\begin{equation}
M=n\langle\mu\rangle L(\langle\mu\rangle H/k_BT)\label{eq:Wohlfarth}
\end{equation}
where $n$ is the number of clusters per unit volume, $\langle\mu\rangle$ is the average magnetic moment of the clusters, $k_B$ is the Boltzman constant, and $L(x)$ is the Langevin function. Above the blocking temperature ($T_B$), the expansion of Langevin function in powers of $H$ gives
\begin{equation}
\chi_1=n\langle\mu\rangle^2/3k_BT=P_1/T\label{eq:Wohlfarth1}
\end{equation}
and
\begin{equation}
\chi_3=(n\langle\mu\rangle/45)(\langle\mu\rangle/k_BT)^3=P_3/T\label{eq:Wohlfarth2}
\end{equation}
The equation \ref{eq:Wohlfarth1} and \ref{eq:Wohlfarth2} show that for superparamagnetic clusters, above $T_B$, $\chi_1$ and $\chi_3$ varies as a linear function of $T^{-1}$ and $T^{-3}$ respectively. The figure \ref{fig: superpara} shows the $\chi^{'}_1$ versus $T^{-1}$ and its inset shows the $\chi^{'}_3$ versus $T^{-3}$ above $T_B$. $\chi^{'}_1$ and $\chi^{'}_3$ curves show nonlinearity up to 185~K (above $T^{-1}\approx5.41\times10^{-3}$~K$^{-1}$ and $T^{-3}\approx1.58\times10^{-7}$~K$^{-3}$) and thereafter exhibit a linear  behavior. The presence of curvature between T$_B$ and 185~K is possibly due to a large variation in the size of ferromagnetic-clusters. 
In the linear region, the ratio of the fitting  parameter $P_3$ and $P_1$ is used to estimate average value of $\langle\mu\rangle$, which comes around $1.83\times10^5~\mu_B$ where $\mu_B$ is the effective Bohr magneton. Such a large value of $\langle\mu\rangle$ is generally observed in superparamagnet clusters. This is because the cluster consists of a large number of atomic spins each having the magnetic moment of few $\mu_B$ (while normal paramagnet only have the atomic spins). Since $\langle\mu\rangle$=$M_SV$ where $M_S$ is the saturation magnetization and $V$ is the volume of cluster, assuming the clusters to be spherical, the average size of the clusters comes around 15~nm. A large variation in the cluster size is expected from the average value. The average size of the magnetic clusters is much smaller than the crystallite size ($\approx 85$~nm calculated from the X-ray diffraction) which indicates that each crystallite may contain a number of ferromagnetic-clusters. Combining this with the results of low temperature isothermal magnetization of section \ref{subsubsection:isothermal magnetization} and frequency dependence of third harmonic of ac susceptibility of section \ref{subsubsection:Nature of magnetically ordered state}, we infer that each crystallite of the system consists of percolating ferromagnetic-clusters coexisting along with relative smaller antiferromagnetic-clusters.

\begin{figure} [!t]
\begin{centering}
\includegraphics[width=0.8\columnwidth]{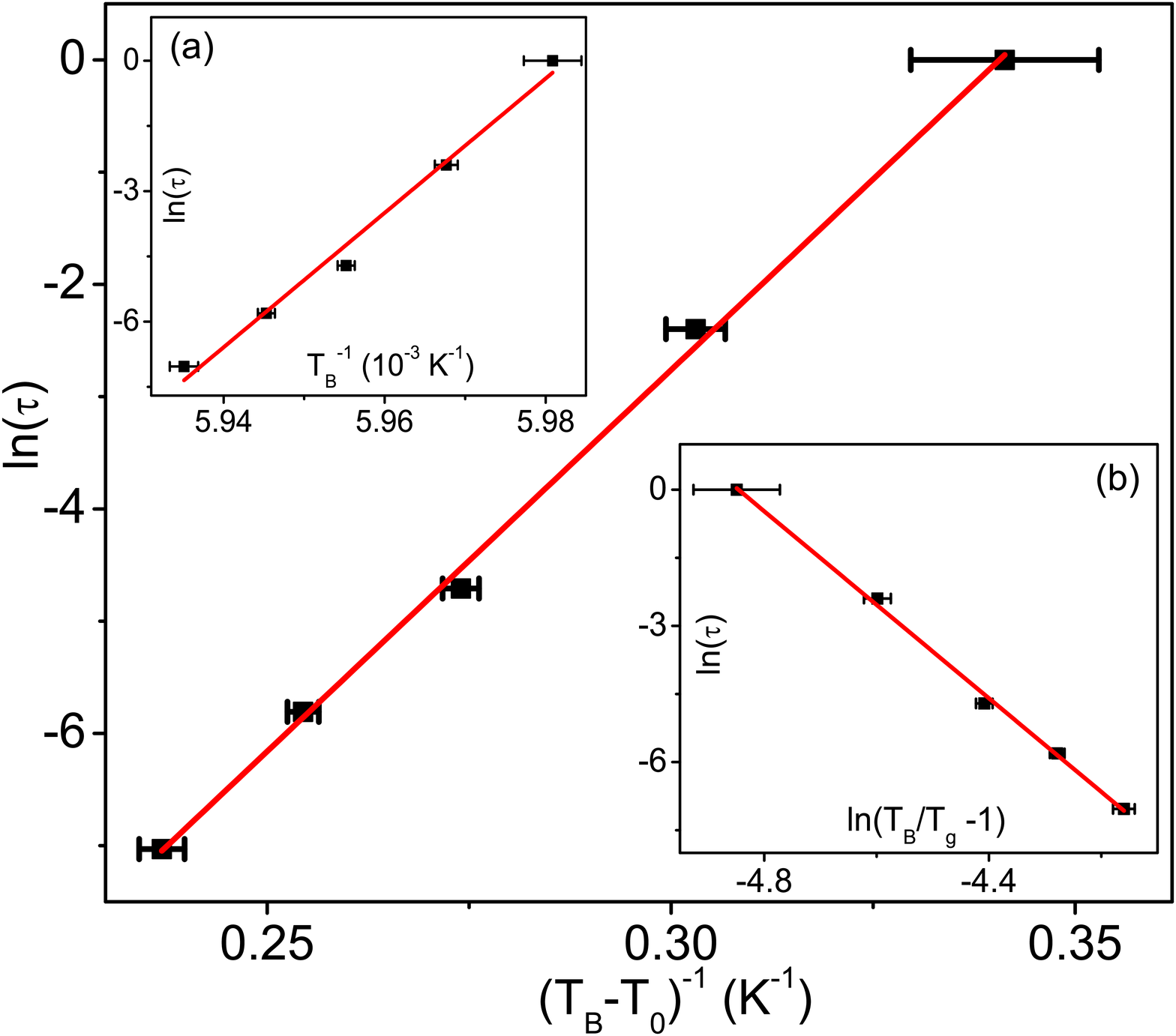}
\par\end{centering}
\caption{Variation of relaxation time ($\tau$) with blocking temperature (T$_B$) plotted as ln$\tau$ versus ($T_B-T_0)^{-1}$. The solid line represents the fitting of Vogel-Fulcher law (equation \ref{eq:VG1}). The inset (a) shows the ln$\tau$ versus T$_B^{-1}$ and the solid line is the fitting of N\'eel-Arrhennius law. The inset (b) shows the ln$\tau$ versus ln$(T_B/T_g -1)$ and the solid line is the fitting of scaling law (equation \ref{eq:dscaling1}).} \label{fig:scaling}
\end{figure}

\subsubsection{Inter-cluster interaction}
The ferromagnetic-clusters coexisting with the non-ferromagnetic phases may interact with each other directly through dipole-dipole interaction or via non-ferromagnetic matrix through exchange interactions.\cite{Bedanta} The degree of inter-cluster interaction and their effect on
fluctuation dynamics is studied by fitting the frequency dependence of $T_B$ with N\'eel-Arrhennius, Vogel-Fulcher, and scaling law.\cite{Binder} To perform these fittings we need T$_B$ and  relaxation time ($\tau$) for different measurement frequencies; $\tau=\nu^{-1}$ while T$_B$ is obtained by the GaussAmp fitting of the corresponding peak in $\chi^{'}_1$ (see figure \ref{fig: ACR}) in the temperature range of approximately 160-174~K.  The results of these fittings are shown in figure \ref{fig:scaling}. For an ensemble of non-interacting superparamagnets, the relaxation time $\tau$ follows the N\'eel-Arrhennius  law\cite{Binder}
\begin{equation}
\tau=\tau_0 \textrm{exp}\left(\frac{E_a}{k_BT}\right)\label{eq:Neel}
\end{equation}
where $E_a$ is the average anisotropy energy barrier, $\tau_0$ is the time constant corresponding to characteristic attempt frequency, and $k_B$ is the Boltzman constant. The experimentally observed $\tau_0$ values for non-interacting superparamagnets are in the range of $10^{-8}-10^{-13}$~s.\cite{Dormann1} The inset (a) of figure \ref{fig:scaling} show the fitting of equation \ref{eq:Neel} to ln$\tau$ versus $T_B^{-1}$ data which gives $\tau_0\approx10^{-402}$~s and $E_a$=13.31~eV. The fitting of N\'eel-Arrhennius law yields un-physical values which rule out the possibility of non-interacting dynamics and hint the presence of cooperative dynamics due to inter-cluster interaction. The dynamics of interacting superparamagnets is described by Vogel-Fulcher law\cite{Binder}
\begin{equation}
\tau=\tau_0 \textrm{exp}\left(\frac{E_a}{k_B(T-T_0)}\right)\label{eq:VG}
\end{equation}
where the temperature $T_0$, which has a value between zero and $T_B$ is often related to the strength of inter-cluster interaction.  The parameters T$_0$, $\tau_0$, and E$_a$/k$_B$ are obtained by fitting of equation \ref{eq:VG} to $\tau$ versus $T_B$, but this method of fitting suffers from unequal weightage given to some experimental data points. In order to fit equation \ref{eq:VG} with nearly equal weight to all data points, we rewrite equation \ref{eq:VG} as
\begin{equation}
\textrm{ln}\tau=\textrm{ln}\tau_0 + \left(\frac{E_a}{k_B(T-T_0)}\right)\label{eq:VG1}
\end{equation}
Now taking T$_0$ value obtained from fitting of equation \ref{eq:VG} to $\tau$ versus $T_B$ data as starting point, T$_0$ is varied in step of 0.1~K and equation \ref{eq:VG1} is fitted to the data of figure \ref{fig:scaling}. The best fit gives correct value of T$_0$. We get $T_0$=164.3(1)~K, $\tau_0=1.1(6)\times$10$^{-10}$~s, and $E_a/k_B$=70(2)~K. The $\tau_0$ value obtained from the Vogel-Fulcher fitting is orders of magnitude larger than the spin-flip time of atomic magnetic moments ($\sim10^{-13}$~s). This strongly supports that the fluctuating entities are spin-clusters with a significant inter-cluster interaction among them.
Strong inter-cluster interactions can give rise to spin-glass like cooperative freezing, and in this case, the frequency dependence of peak in $\chi^{'}_1$ is expected to follow the power law divergence of the standard critical slowing down given by dynamic scaling theory\cite{Maydosh, Binder}
\begin{equation}
\tau=\tau_0(T/T_g-1)^{-z\nu^{'}}\label{eq:dscaling}
\end{equation}
where $\tau$ is the dynamical fluctuation time scale corresponding to measurement frequency at the peak temperature of $\chi^{'}_1$, $\tau_0$ is the spin flipping time of the relaxing entities, $T_g$ is the  cluster-glass (or spin-glass) transition temperature in the limit of zero frequency, $z$ is the dynamic scaling exponent, and $\nu^{'}$ is the critical exponent. In the vicinity of cluster-glass transition, the spin cluster correlation length $\xi$ diverges as $\xi\propto(T/T_g-1)^{-\nu^{'}}$ and the dynamic scaling hypothesis relates $\tau$ to $\xi$ as $\tau\sim\xi^{z}$. To fit the data with nearly equal weight, it is convenient to rewrite equation \ref{eq:dscaling} as
\begin{equation}
\textrm{ln}\tau=\textrm{ln}\tau_0 -z\nu^{'}\textrm{ln}(T/T_g-1)\label{eq:dscaling1}
\end{equation}
The inset (b) of figure \ref{fig:scaling} show ln$\tau$ versus ln$(T/T_g -1)$. Starting with $T_g$ value obtained from fitting of equation \ref{eq:dscaling} to $\tau$ versus $T_B$ data, $T_g$ is varied in step of 0.1~K to obtain the best fit of equation \ref{eq:dscaling1} to  the data of inset (b) of figure \ref{fig:scaling}. This gives $T_g$=165.9(1)~K, $\tau_0\sim10^{-21}$~s, and $z\nu^{'}$=10.3(3). The value of exponent $z\nu^{'}$ is somewhat higher than that observed in case of spin-glasses (2-10) and $\tau_0$ is orders of magnitude smaller than the values reported for cluster-glasses ($10^{-6}$-$10^{-10}$~s) and spin-glasses ($10^{-11}$-$10^{-13}$~s). The value of $\tau_0$ is even smaller than the spin-flip time of a single atom ($\sim10^{-13}$~s), which is un-physical, and this indicates that the spin-cluster dynamics in the system does not exhibit the critical slowing down on approaching $T_g$ as expected from the dynamic scaling theory. Thus, it can be concluded that the inter-cluster interactions among the ferromagnetic-clusters are significant, but not strong enough to cause a spin-glass like transition.

\begin{figure} [!t]
\begin{centering}
\includegraphics[width=0.7\columnwidth]{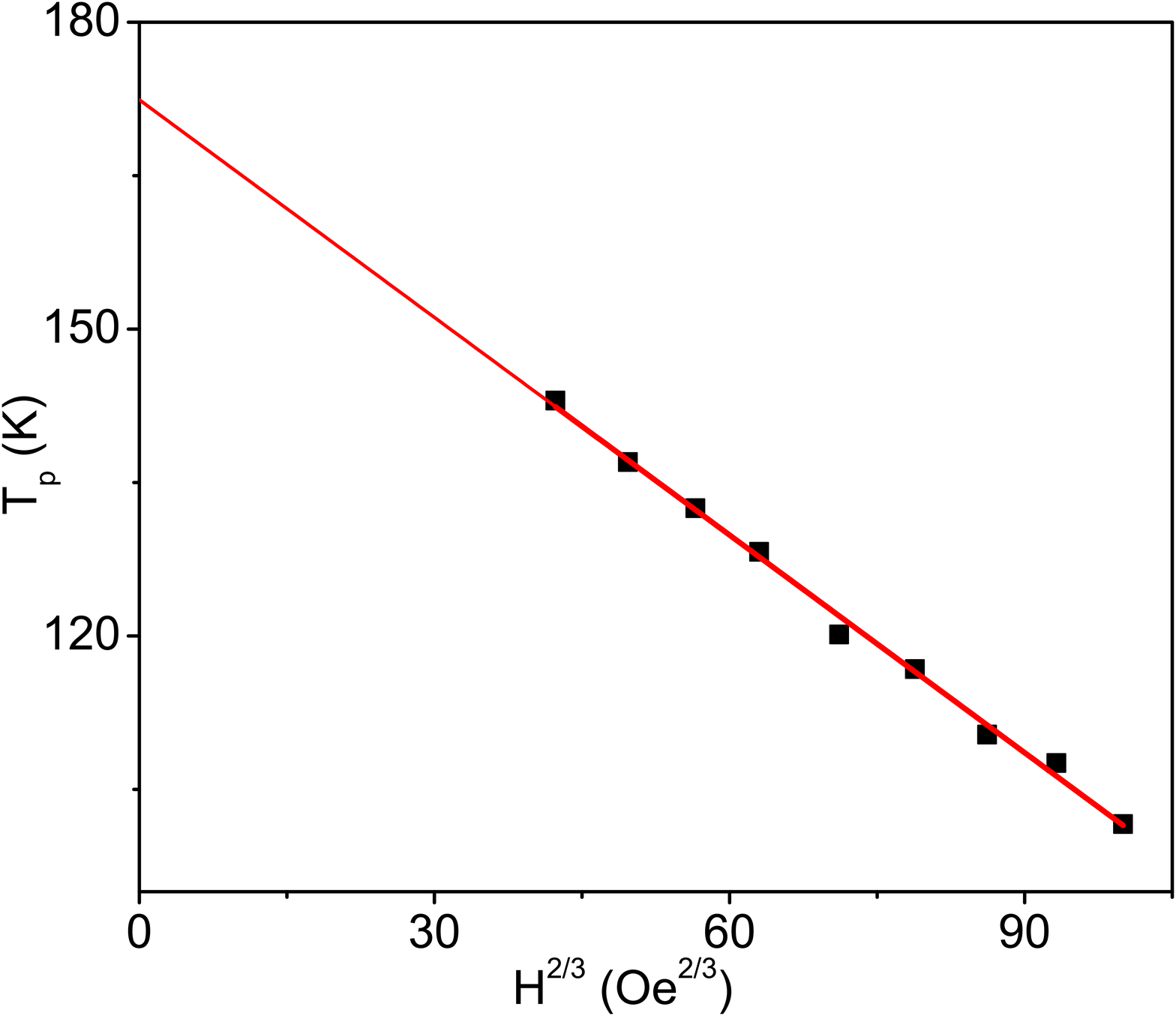}
\par\end{centering}
\caption{Field dependence of the peak in ZFC magnetization. The straight line shows the fitting of AT line to data.} \label{fig:ATline}
\end{figure}

\subsection{Further discussions}
The presence of significant inter-cluster interaction can also be reaffirmed from the field dependence of the peak temperature (T$_p$) in ZFC magnetization curves. For Ising spin-glass, mean field theory of spin-glass predicts a critical de Almeida-Thouless (AT) line in $H$-$T$ space which marks the spin-glass phase transition.\cite{Almeida} Above AT line the large field destroys the frozen spin state. The spin-glass transition temperature corresponds to the peak in ZFC magnetization (T$_p$) and the AT line predicts that T$_p\propto H^{2/3}$.  The AT line like field dependence of $T_p$ is not unique to spin or cluster -glass transition but it has been also observed in some of interacting super-paramagnets which otherwise undergo a progressive thermal blocking.\cite{Pramanik, Wenger} In figure \ref{fig:ATline} we have plotted the field dependence of T$_p$ which fits well with the AT line giving zero field spin-glass transition temperature ($T_g$) of 172(1)~K. Since our ac susceptibility measurements have ruled out the possibility of a spin or cluster -glass like freezing, the existence of AT line in $H$-$T$ space clearly indicates the presence of a significant inter-cluster interaction in the system. The inter-cluster interaction can originate from different type of magnetic interactions and the strength of these interactions generally depends on the packing density of ferromagnetic-clusters.
These magnetic interactions include the long range dipole-dipole interaction among the ferromagnetic-clusters along with the possibilities of exchange, tunneling exchange and superexchange interactions.\cite{Bedanta}

The absence of spin-glass phase in La$_{0.5}$Ba$_{0.5}$CoO$_3$ is in contrast with La$_{0.5}$Sr$_{0.5}$CoO$_3$ where spin-glass phase coexist along with the percolating ferromagnetic-clusters.\cite{Tang, Samal1} 
The doping at A site of LaCoO$_3$ with 50\% Ba or Sr gives same hole concentration, and therefore, the observed discrepancy in magnetically ordered state of La$_{0.5}$Ba$_{0.5}$CoO$_3$ and La$_{0.5}$Sr$_{0.5}$CoO$_3$ can be only due to difference in local lattice distortions caused by the difference in ionic radii of Ba$^{2+}$ and Sr$^{2+}$. The elastic neutron scattering of the Sr and Ba doped LaCoO$_3$ indicate the existence of an incommensurate magnetic ordering with antiferromagnetic correlations along with the ferromagnetic-clusters.\cite{Phelan2, Yu1} While for La$_{1-x}$Ba$_{x}$CoO$_3$ the strength of incommensurate state increases and becomes commensurate on increasing $x$, for La$_{1-x}$Sr$_{x}$CoO$_3$ incommensurate state strengthens but remains incommensurate on increasing $x$.\cite{Yu1} This is because the enhanced local randomness in La$_{1-x}$Ba$_{x}$CoO$_3$ due to larger ionic radii of Ba$^{2+}$ favour the growth of antiferromagnetic ordered phases. The existence of an competing incommensurate-antiferromagnetic-ordering along with ferromagnetic-clusters possibly gives the spin-glass phase in La$_{0.5}$Sr$_{0.5}$CoO$_3$. In La$_{0.5}$Ba$_{0.5}$CoO$_3$ both the ferromagnetic and antiferromagnetic ordering are commensurate resulting in coexisting antiferromagnetic and ferromagnetic -clusters. The absence of spin-glass phase in  La$_{0.5}$Ba$_{0.5}$CoO$_3$ suggests that the spin-glass phase in La$_{0.5}$Sr$_{0.5}$CoO$_3$ is associated with the presence of incommensurate-antiferromagnetic-ordering in the non-ferromagnetic phases.  

\section{Conclusions}
In conclusion, we have performed a comprehensive set of dc magnetization, linear and non-linear ac susceptibility measurements to understand the magnetic state of the hole doped disordered cobaltite La$_{0.5}$Ba$_{0.5}$CoO$_3$. The results of isothermal magnetization suggest that the magnetically ordered state of the system consists of percolating ferromagnetic-clusters coexisting along with the non-ferromagnetic phases. 
The percolating ferromagnetic-clusters possibly start a magnetic ordering around 201.5(5)~K. The frequency dependence of the third harmonic of ac susceptibility suggests that the non-ferromagnetic phases mainly consist of antiferromagnetic-clusters whose sizes are smaller than that of ferromagnetic-clusters.

Below $T_C$ the system exhibits thermomagnetic irreversibility and frequency dependence in the peak of ac susceptibility which suggest the presence of spin-glass, cluster-glass, or superparamagnetic phases. The absence of field divergence in the peak of third harmonic of ac susceptibility and absence of ZFC memory rule out the existence of spin or cluster -glass phase and  suggest that the observed spin-dynamics is possibly due to superparamagnet like thermal blocking of ferromagnetic-clusters. This is in sharp contrast to La$_{0.5}$Sr$_{0.5}$CoO$_3$ where the spin-glass phase coexist along with the ferromagnetic-clusters. Our analysis suggests that the existence of spin-glass phase is associated with the presence of incommensurate antiferromagnetic ordering in the non-ferromagnetic phases which in turn is determined by the degree of local lattice distortion caused by the doping of divalent ion. The superparamagnetic behavior of ferromagnetic-clusters in La$_{0.5}$Ba$_{0.5}$CoO$_3$ is further confirmed by Wohlfarth's model of superparamagnetism. The analysis of frequency dependence in the peak of ac susceptibility by N\'eel-Arrhennius, Vogel-Fulcher, and scaling law suggest the existence of significant inter-cluster interaction among the ferromagnetic-clusters which is further confirmed by the existence of an AT line in the H-T space.

\ack
We are thankful to P. Chaddah for useful discussions. We gratefully acknowledge M.~Gupta for XRD and N. P. Lalla for EDAX measurements.

\end{document}